\documentclass[aps,prd,twocolumn,superscriptaddress,footinbib]{revtex4-2}

\pdfoutput=1
\usepackage{graphicx}
\usepackage{subcaption}
\graphicspath{{./picture/}}
\usepackage{empheq}
\usepackage{amssymb}
\usepackage{mathtools}
\usepackage{commath}
\usepackage{verbatim}
\usepackage{xcolor}
\usepackage{soul}
\usepackage{booktabs}
\usepackage{amsmath}
\usepackage{hhline}
\usepackage{url}
\usepackage{ragged2e}
\usepackage{caption}
\DeclareCaptionFormat{revtex}{#1#2\justifying{#3}}
\usepackage{epsfig}
\usepackage{mathrsfs}
\usepackage{nicefrac} 
\usepackage{upgreek}
\usepackage{bbm}
\usepackage{psfrag}

\usepackage{hyperref}

\usepackage[inline,shortlabels]{enumitem}

\newcommand{\g}{\mathsf{g}}

\newcommand{\V}{\mathcal{V}}
\newcommand{\Vend}{\mathcal{V}_{\rm end}}
\newcommand{\x}{\textbf{x}}

\newcommand{\T}{{\rm T}}

\newcommand{\rhoend}{\rho_{\rm end}}
\newcommand{\rhoreh}{\rho_{\rm re}}

\newcommand{\wrehbar}{\bar{w}_{\rm re}}

\newcommand{\Nreh}{N_{\rm re}}
\newcommand{\X}{X}

\newcommand{\M}{M}
\usepackage{slashed}				
\newcommand{\Treh}{T_{\rm re}}
\newcommand{\GG}{\Gamma}
\newcommand{\Nk}{N_{\rm k}}
\renewcommand{\c}{{\rm c}}
\newcommand{\DD}{D}
\newcommand{\n}{ j}
\newcommand{\epsilonvarM}{\frac{m_\varphi}{M_{pl}}}
\newcommand{\epsilonvar}{\frac{m_\varphi}{\varphi_{\rm end}}}
\newcommand{\vv}{\mathsf{v}}
\newcommand{\xx}{a}

\newcommand{\sigmar}{\sigma_r}
\newcommand{\sigmans}{\sigma_{n_s}}
\newcommand{\nsbar}{\bar{n}_s}
\newcommand{\rbar}{\bar{r}}
\newcommand{\nn}{{ 1}}

\newcommand{\sigmax}{\sigma_\x}
\newcommand{\lambdaphi }{\lambda}
\newcommand{\EFTscale}{\Lambda}

\begin{document}


\title{AliCPT Sensitivity to Cosmic Reheating}


\author{Yang Liu}
\email{liuy92@ihep.ac.cn}
\affiliation{Key Laboratory of Particle Astrophysics, Institute of High Energy Physics (IHEP), Chinese Academy of
	Sciences, 19B Yuquan Road, Beijing, China}

\author{Lei Ming}
\email[Corresponding author: ]{minglei@scnu.edu.cn}
\affiliation{Key Laboratory of Atomic and Subatomic Structure and Quantum Control (Ministry of Education), Guangdong Basic Research Center of Excellence for Structure and Fundamental Interactions of Matter, School of Physics, South China Normal University, Guangzhou 510006, China} 
\affiliation {Guangdong Provincial Key Laboratory of Quantum Engineering and Quantum Materials, Guangdong-Hong Kong Joint Laboratory of Quantum Matter, South China Normal University, Guangzhou 510006, China}

\author{Marco Drewes}
\email{marco.drewes@uclouvain.be}
\affiliation{Centre for Cosmology, Particle Physics and Phenomenology, Universit\'{e} catholique de Louvain, Louvain-la-Neuve B-1348, Belgium}
\affiliation{Physik–Department, Technische Universität München, James Franck Straße 1, D-85748 Garching, Germany}

\author{Hong Li}
\email{hongli@ihep.ac.cn}
\affiliation{Key Laboratory of Particle Astrophysics, Institute of High Energy Physics (IHEP), Chinese Academy of
	Sciences, 19B Yuquan Road, Beijing, China}


\begin{abstract}
We present the first assessment of the Ali Cosmic Microwave Background Polarization Telescope's (AliCPT) sensitivity to the reheating epoch after cosmic inflation, based on its  ability to detect primordial gravitational waves.  We consider three models of inflation, an $\alpha$-attractor T-model, RGI inflation and QCD-driven warm inflation. Assuming a fiducial value of $r=0.01$, we find that AliCPT-1, in its fully loaded focal plane detector configuration and combined with Planck, can provide measurements of the order of magnitude of the reheating temperature with an accuracy around $10\%$. For QCD-driven warm inflation this can be translated into a constraint on the inflaton coupling to gluons, which can be probed independently in axion search experiments. Our results constitute the first demonstration of AliCPT's ability to probe the initial temperature of the hot big bang and the microphysical parameter connecting cosmic inflation and particle physics.
\end{abstract}

\maketitle

\begin{footnotesize}
\tableofcontents
\end{footnotesize}
\section{Introduction}

Observations of the Cosmic Microwave Background (CMB) have not only played a key role in establishing the concordance model of cosmology ($\Lambda$CDM) \cite{WMAP:2003elm}, they also confirm that the fundamental laws of nature known from experiments on Earth hold in the most distant regions of the observable universe and at times predating the first sunrise by almost ten billion years. They in particular confirm that the early universe was filled with a hot plasma 
at a temperature $T$ hotter than the surface of the Sun \footnote{Indirectly CMB observations indicate that $T$ was at least a thousand times hotter than the Sun’s core, c.f.~\cite{deSalas:2015glj} and references therein. This conclusion can also be drawn from the light element abundances in the intergalactic medium \cite{Cyburt:2015mya} alone.}. 
However, neither the $\Lambda$CDM model 
nor our most fundamental established theories of Nature, the Standard Model of particle physics (SM) and General Relativity,
can explain the initial conditions of this \emph{hot big bang}---including the large scale homogeneity and isotropy of the cosmos, its small overall spacial curvature, the origin of the tiny temperature fluctuations that formed the seeds for galaxy formation,
and the initial temperature $\Treh$
at the onset of the radiation dominated epoch.

Fluctuations in the CMB are believed to contain information about time predating this epoch, potentially giving insights into the mechanism that set these initial conditions. 
The possibly most promising solution that can explain all of the aforementioned problems is offered by \emph{cosmic inflation} \cite{Starobinsky:1980te,Guth:1980zm,Linde:1981mu}, i.e., the idea that the universe underwent a period of accelerated expansion at very early times.
However, it is not clear what mechanism drove cosmic inflation, and how it is implemented in a fundamental microphysical theory of Nature beyond the SM. 
Many of the proposed explanations can be parameterised by a single scalar field $\varphi$, dubbed inflaton, with a potential $\V(\varphi)$, see e.g.~\cite{Martin:2013tda} for a partial list. Unfortunately observational data currently is not sufficient to single out a particular choice of $\V$.
Even less is known about the interactions between $\varphi$ and other fields---which would be crucial to understand how a given inflationary model, characterized by $\V$, is embedded in a theory of particle physics, and eventually connected to the SM. 

If inflation happened, it  diluted all particle populations in the universe, implying that the energy density $\rhoend$ at the end of inflation was dominated by the 
contribution from the zero-mode of $\varphi$, 
$\rho_\varphi \simeq \frac{1}{2}\dot\varphi^2 + \V$.
Converting the energy density $\rho_\varphi$ into other particles necessarily took some time, hence inflation was separated from the radiation dominated epoch by a period of 
\emph{cosmic reheating}~\cite{Albrecht:1982mp,Dolgov:1989us,Traschen:1990sw,Shtanov:1994ce,Kofman:1994rk,Boyanovsky:1996sq,Kofman:1997yn} of unknown duration.  
It is well-known that the CMB is sensitive to some properties of this reheating epoch \cite{Lidsey:1995np}, 
in particular its duration in terms of e-folds $\Nreh$ and the averaged equation of state $\wrehbar$, 
from which one can determine $\Treh$ \cite{Martin:2010kz,Adshead:2010mc,Mielczarek:2010ag,Easther:2011yq,Dai:2014jja}. 
Since reheating is driven by the inflaton's microphysical interactions, this also provides information on the microphysics of reheating \cite{Drewes:2015coa,Martin:2016iqo}, and in particular the inflaton coupling $\g$ to other particles \cite{Ueno:2016dim,Drewes:2017fmn,Drewes:2019rxn,Ellis:2021kad}, shedding light on the connection between inflation and particle physics.
While present data from the combined 
  Planck 2018 and BICEP/Keck (Planck+BK) observations \cite{BICEP:2021xfz}
already provide some information about reheating \cite{Martin:2014nya,Martin:2016oyk}, future observations can be able to provide a measurement \footnote{By a \emph{measurement} we mean the ability to impose both an upper and a lower bound on a quantity that is stronger than the information available prior to the observation, see e.g.~\cite{Drewes:2023bbs} for a concrete discussion.} of $\Treh$ \cite{Drewes:2022nhu}.

The observational situation is expected to improve dramatically in the future. 
The main properties of the CMB from which information on $\Treh$ and $\g$ can be extracted can be parametrized in terms of the amplitude of scalar perturbations $A_s$, the spectral index $n_s$ and the tensor-to-scalar ratio $r$.
JAXA's LiteBIRD satellite \cite{LiteBIRD:2022cnt} 
is expected reduce the error on $r$ to $\sigma_r \lesssim 10^{-3}$,  enabling measurements of both $\Treh$ and $\g$ in given models \cite{Drewes:2022nhu}.
An even better sensitivity to reheating could be achieved with a ground-based program like CMB-S4 \cite{CMB-S4:2020lpa}, combining both would further improve the measurements of $\Treh$ and $\g$ due to their complementary sensitivity to $A_s$ and $n_s$ \cite{Drewes:2023bbs}.
Before that, significant improvements can already be expected from the Simons Observatory \cite{SimonsObservatory:2018koc}, 
upgrades at the South Pole Observatory \cite{Moncelsi:2020ppj}
and the Chinese Ali Cosmic Microwave Background Polarization Telescopes (AliCPT) observatory in Tibet \cite{Li:2017drr,Li:2018rwc}.

In the present work we perform the first estimate of AliCPT's potential to gain information on the reheating temperature $\Treh$ and the microphysical coupling constant $\g$ that connects the inflaton to other particles.  
Our study is based on the configuration known as AliCPT-1, which is its first CMB telescope in Ali station. AliCPT is currently the only high-altitude, ground-based experiment in the Northern Hemisphere dedicated to the search for primordial B-mode polarization in the CMB. In light of funding uncertainties affecting other ambitious projects---particularly CMB-S4---it may play an increasingly important role, 
in particular with the planned further expansion of its detector array beyond the AliCPT-1 phase.

AliCPT-1 is the first CMB polarization telescope deployed at a high-altitude site in Tibet’s Ali region, operating at 5,250 meters above sea level. Equipped with a 72 cm aperture and a two-lens refracting polarimeter, it observes in two frequency bands (90 GHz and 150 GHz) to measure CMB polarization in the Northern Hemisphere with high sensitivity. The telescope’s focal plane can house 19 detector modules, enabling a total of up to 32,376 detectors. The primary scientific objectives of AliCPT-1 is to detect primordial gravitational waves (B-modes of the CMB) to investigate the first moments of cosmic history, 
providing a new observational window for these studies in the northern sky. Notably, the Northern Hemisphere contains low-foreground emission regions, as identified in the Planck 353 GHz map \cite{Planck:2018yye},  making it an ideal target for deep scans to detect primordial signals. The near future upgrades plan to incorporate low-frequency channels (e.g., 27 GHz and 40 GHz) to further improve foreground separation.
In this work, we simulate five years of AliCPT-1 observations to estimate its constraints on inflation model parameters and derive bounds on reheating model parameters.

\begin{figure}
	\centering
	\includegraphics[width=0.9\linewidth]{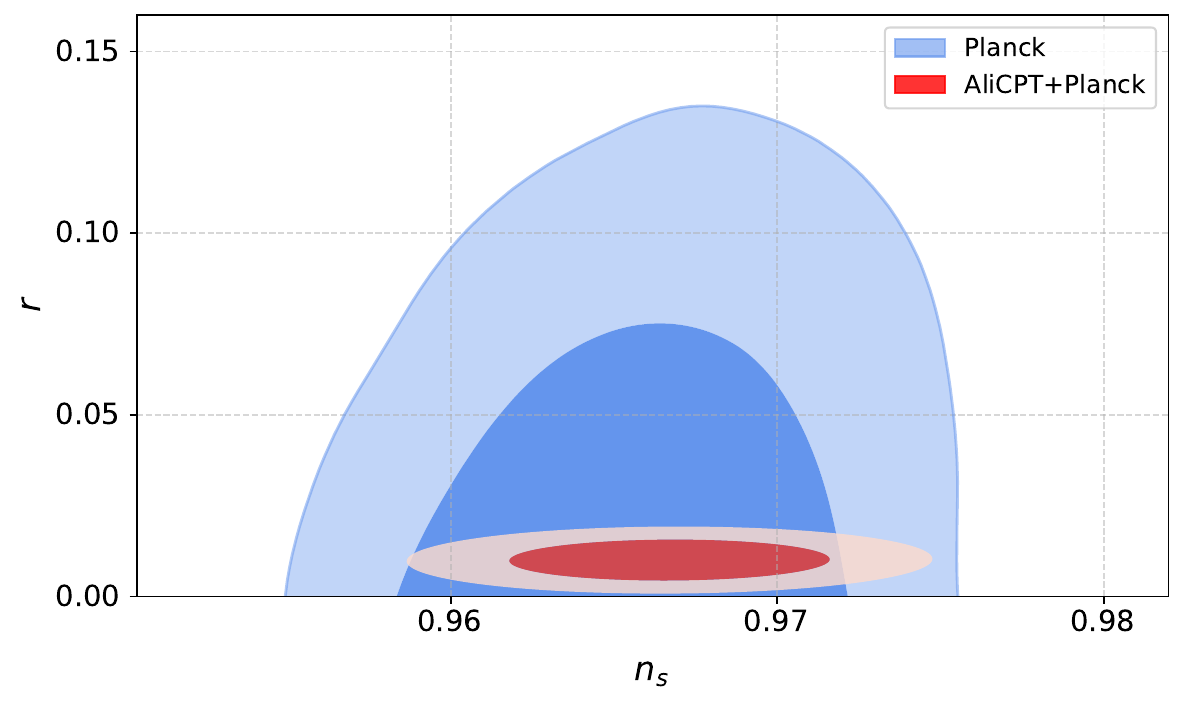}
	\caption{ 
    The red ellipses indicate the 68\% and 95\% CL regions corresponding to the principal observational benchmark \ref{it:A} used throughout this work, cf.~Table \ref{PrincipalBenchmark}. Details are given in Sec.~\ref{sec:forecasts}.
    For comparison, we plot the corresponding regions from the Planck's $MCMC$ chains \cite{Planck:2018vyg}. 
        } 
	\label{fig:ns_vs_r}
\end{figure}

\section{Theoretical background}\label{sec:theory}

We define the beginning of reheating as the moment when the growth of the scale factor $a$ stops accelerating ($\ddot{a}\leq 0$), and its end as the moment when the energy density $\rho_R$ of the radiation bath of particles produced during reheating  exceeds the inflaton energy density $\rho_\varphi$. 
There are several interesting physical questions concerning the reheating epoch that can be asked in the context of next generation observations,
\begin{enumerate*}[I)]
\item \label{it:expansion} How much can we learn about the cosmic \emph{expansion history}, i.e., the time evolution of the scale factor $a$?
\item \label{it:thermal} How much can we learn about the \emph{thermal history} of the universe, i.e., the time evolution of the temperature $T$?
\item \label{it:micro} How much can we learn about the microphysics, i.e., the connection between models of inflation and particle physics theories within which they may be embedded?
\end{enumerate*}

Addressing question \ref{it:expansion} within a given model of inflation relies only on minimal assumptions. 
The equation of state $w$ during reheating in general differs from that during both inflation and the radiation dominated epoch. 
This modifies the redshifting of cosmological perturbation during the reheating period. As a result, the relation between the perturbations generated during inflation and the temperature fluctuations that are observed in the CMB today directly depends on the expansion history during reheating. 
The first  objective of the present work is to assess AliCPT-1's sensitivity to this impact, which can be parameterised by the  number $\Nreh$ of $e$-folds  during reheating (defined as the logarithm of the change in the scale factor $a$)
 and the averaged equation of state $\wrehbar = \frac{1}{N_{\rm re}}\int_0^{N_{\rm re}} w(N) dN$.

Constraining $\Nreh$ and $\wrehbar$ also 
imposes bounds on $\Treh$ \footnote{In principle the thermal gravitational wave background \cite{Ghiglieri:2015nfa} offers a more direct probe of $\Treh$ \cite{Ghiglieri:2020mhm,Ringwald:2020ist},  but it is very challenging in practice \cite{Drewes:2023oxg} to observe it unless processes other than pure thermal emission produce a signature from reheating (cf.~\cite{Caprini:2018mtu,Roshan:2024qnv}).}, see \eqref{Tre} below, and therefore partially addresses question \ref{it:thermal} concerning the thermal history. 
However, in standard inflation, 
the complete thermal history during reheating  is not accessible in this manner, 
as the impact of the radiation on the expansion history was negligible ($\rho_R<\rho_\varphi$ by definition during reheating).  
The thermal history can be reconstructed with knowledge about the inflaton couplings to other fields, linking it  to question \ref{it:micro}.
A second objective of this work is to identify scenarios in which 
these inflaton couplings can be measured and the thermal history during reheating may reconstructed, 
i.e., questions \ref{it:thermal} and \ref{it:micro} can be addressed with AliCPT.

While the CMB in principle comprises a tremendous wealth of data, the relevant information concerning inflation and reheating can be expressed in terms of a handful of numbers, in particular the amplitude of scalar perturbations $A_s$, the scalar spectral index $n_s$ and the scalar-to-tensor ratio $r$ \footnote{In the future the running of $n_s$ or non-Gaussianities could provide additional information.}. 
This imposes a fundamental limitation on the number of microphysical parameters that can be constrained from CMB data,  
which represents a challenge to the empirical testability of inflation for two distinct reasons.
\begin{enumerate}[1)]
\item \label{challenge1} Firstly, for standard single field inflation, observational data exclude most simple  potentials with a single parameter $\vv$, in particular monomial shapes $\V \propto \vv\varphi^n$. 
Potentials with more parameters remain viable, but the limited information contained in the three numbers $(n_s, A_s,r)$ is generally insufficient to reconstruct all of them. Hence, it is not only impossible to single out a specific model from the vast menu of scenarios, but even the perspectives to constrain the parameter space of a given model are very limited.
\item \label{challenge2} Secondly, in most particle physics models the inflaton does not directly (or not exclusively) interact with the known particles of the SM, introducing additional unknown parameters in the form of couplings to mediator fields. 
If the backreaction of the produced particles on the rate $\GG$ of their production is significant (implying that it affects the efficiency and hence the duration of the reheating process), 
then $\Nreh$ generally depends on a considerable number of unknown parameters related to the properties of other hypothetical particles that act as mediators in the reheating process. 
Information on three numbers $(n_s, A_s,r)$ is insufficient to constrain this potentially large set of parameters (in addition to those in $\V$).
While this does not impose a further obstacle to measuring $\Treh$ or the parameters in $\V$ in addition to the previous point \ref{challenge1}, 
it further diminishes the ability to measure the inflaton coupling $\g$ or other microphysical parameters.
\end{enumerate}

We take two different approaches to mitigate the impact of these fundamental limitations.
\begin{enumerate}[a)]
\item \label{it:approachA} 
\emph{Avoid backreaction}. One possibility is to consider scenarios where backreaction during reheating is negligible, so that $\Nreh$ can be related to a single inflaton coupling $\g$. Generally this requires comparable small values of $\g$ (see appendix \ref{sec:appendixB}), 
but for the Yukawa interaction considered in Sec.~\ref{sec:cold} 
it can be justified even for large values of the coupling constant.
We present results based on this approach in Sec.~\ref{sec:ColdModel}. 
\item \label{it:approachB} 
\emph{Minimal models.}
Another possibility is to focus on minimal extensions of the SM in which reheating is driven by a direct coupling between the inflaton and SM fields, eliminating the uncertainty coming from the properties of unknown mediators. 
In Sec.~\ref{sec:warm}  we consider such a scenario in which 
the total number of relevant model parameters remains small enough to be testable.
We present results for this minimal model in Sec.~\ref{sec:WarmModel}.
\end{enumerate}

For the purpose of the following discussion we assume that the evolution of $\varphi$ and the radiation energy density $\rho_R$ in the relevant regimes can effectively be described by an equation of motion of the form \footnote{See appendix C in \cite{Drewes:2019rxn} for a discussion on the range of validity of \eqref{EOM}.}
\begin{subequations}\label{EOM}
\begin{eqnarray}
    \ddot{\varphi} + (3H + \GG)\dot{\varphi} + \partial_\varphi \V, &=&0 \label{EOM1}\\
    \dot{\rho}_R + 4 H \rho_R &=& \GG \dot{\phi}^2.\label{EOM2}
\end{eqnarray}
\end{subequations}

\subsection{Standard (cold) single-field inflation}\label{sec:cold}

The spectrum of primordial perturbations produced during inflation 
and the inflaton field value at the end of inflation $\varphi_{\rm end}$ are both determined by the inflationary model, i.e., by specifying the inflaton potential $\V$. 
This also fixes the energy density at the end of inflation $\rhoend\simeq \frac{4}{3}\V(\varphi_{\rm end}) \equiv \frac{4}{3}\Vend$.  
Leaving aside late time and foreground effects and assuming a standard cosmic history after reheating, the observable spectrum of CMB perturbations for given $\V$ could be predicted once the expansion history during reheating was known. The current uncertainty in this prediction from reheating can be parameterised in terms of $\wrehbar$ and $\Nreh$.

\subsubsection{Constraining the reheating temperature $\Treh$}\label{sec:Treh}

At leading order in the slow roll parameters \footnote{Next-to-leading order corrections in the slow roll parameters can be neglected in this context, see appendix B in \cite{Drewes:2023bbs}. Other uncertainties, such as the mild dependence on $g_*$, are also sub-dominant, see Sec.~2 in \cite{Drewes:2019rxn}.}, 
$\epsilon=\left(\partial_\varphi \V/\V\right)^2 M^2_{pl}/2$ and 
$\eta=M^2_{pl}\partial^2_\varphi \V/\V$, reheating begins when $\epsilon = 1$, and it ends when the rate of particle production $\GG$ exceeds the Hubble rate $H =\dot{a}/a$.
The equation of state $\wrehbar$ is determined by specification of $\V$ because 
the total energy density $\rho$ is still dominated by $\rho_\varphi$ during reheating (by definition), leaving $\Nreh$ as the sole relevant quantity that is not fixed by the choice of $\V$. 
With the redshifting relation $\rho \propto \exp(- 3N(1 + w))$,
$\Nreh$ can be related to the energy density at the end of reheating $\rhoreh= \rhoend\exp(- 3\Nreh(1 + \wrehbar))$.
The latter is often parameterised in terms of an effective reheating temperature defined by $\frac{\pi^2 g_*}{30}T_{\rm re}^4 \equiv \rhoreh$ \footnote{$\Treh$ can only interpreted as a temperature in strict sense if the interactions among the produced particles are fast enough to bring them into local thermal equilibrium on time scales much shorter than $\Nreh$. However, $\Treh$ can be a useful parameter that characterizes typical particle energies even if this is not the case, see~\cite{Berges:2008wm,Mazumdar:2013gya,Harigaya:2013vwa,Harigaya:2014waa,Mukaida:2015ria,Mukaida:2024jiz} and references therein for discussions.}, where $g_*$ denotes the effective number of relativistic degrees of freedom,
\begin{equation}\label{Tre}
	\Treh=\exp\left[-\frac{3(1+\bar{w}_{\rm re})}{4}N_{\rm re}\right]\left(\frac{40 \ \Vend }{g_*\pi^2}\right)^{1/4}.
\end{equation}
The connection between $\Nreh$ and observables $n_s$ and $r$ is well-known, a derivation can e.g.~be found in \cite{Ueno:2016dim}. 
We recapitulate the most important relations in appendix \ref{sec:appendixA}.

\subsubsection{Connection to particle physics}\label{sec:ConnectionParticlePhysics}
The duration of  reheating $\Nreh$ is determined by the efficiency of the dissipative processes that transfer energy from $\varphi$ into the radiation bath formed by the produced particles. 
This efficiency is given by the effective particle production rate $\GG$. 
Using the Friedmann equation $H^2=\rho/(3 M_{pl}^2)$ and the fact that reheating ends when $\GG=H$ one finds
\begin{equation}
	\GG|_{\GG=H}
	=\frac{1}{M_{pl}}\left(\frac{\rhoend}{3}\right)^{1/2} e^{-3(1+\wrehbar) \Nreh/2} 
	\simeq \frac{\Treh^2}{M_{pl}}\frac{\sqrt{g_*}}{3},\label{GammaConstraint}
\end{equation}
where $M_{pl}=2.435\times 10^{18}~{\rm GeV}$ is the reduced Planck mass. 
The transfer of energy from $\varphi$ into other degrees of freedom is driven by microphysical interactions of the inflaton, hence $\GG$ is in principle calculable in terms of particle physics parameters, and any constraint on $\Nreh$ (or likewise $\Treh$) can be translated into information on those parameters by means of the relation \eqref{GammaConstraint}. 

For instance, if reheating is driven by perturbative $1\to 2$ decays, then one generally finds 
\begin{eqnarray}\label{GammaPerturbative}
	\GG = \g^2 m_\varphi /\c,
\end{eqnarray}
where $\c$ is a numerical factor that depends on the type of interaction  
and the inflaton mass $m_\varphi$ is defined by expanding $ \V = \sum_\n\frac{\vv_{\n}}{\n !} \frac{\varphi^\n}{\EFTscale^{\n-4}}  = \frac{1}{2}m_\varphi^2\varphi^2+\frac{g_\phi}{3!}\varphi^3+\frac{\lambdaphi }{4!}\varphi^4 + \ldots$, with $\EFTscale$ an appropriately chosen mass scale.
If \eqref{GammaPerturbative} holds when $\GG=H$
we can obtain constraints on the microphysical coupling $\g$ 
that connects inflation to particle physics 
by plugging \eqref{GammaPerturbative} into \eqref{GammaConstraint},
\begin{eqnarray}\label{CouplinfToTreh}
	\g = \frac{\Treh}{\sqrt{m_\varphi M_{pl}}} \sqrt{\frac{\pi c}{3}} (g_*/10)^{1/4}.
\end{eqnarray}

A conservative estimate of the parameter range where \eqref{CouplinfToTreh} can applied is given by the conditions outlined in appendix \ref{sec:appendixB},
which in 
the plateau models considered in Sec.~\ref{sec:ColdModel} amounts to \cite{Drewes:2019rxn}
\begin{eqnarray}\label{PerturbativityContraintsPleteau}
\lambdaphi 
\ll
3\pi^2  r A_s 
\ , \  
|\g| \ll \left(3\pi^2  r A_s\right)^{1/2}.
\end{eqnarray}
The condition on $\g$ generally limits the range of validity of \eqref{CouplinfToTreh} to very small coupling constants. 
This range can, however, be considerably increased with comparably moderate model assumptions (cf.~Appendix \ref{sec:appendixB}) in scenarios where $\GG$ dominated by a Yukawa coupling to fermions $\psi$ of the form $y\varphi\bar{\psi}\psi$, 
corresponding to $(\g,c)=(y,  8\pi)$ in \eqref{GammaPerturbative},
in which case \eqref{CouplinfToTreh} yields
\begin{eqnarray}\label{YukawaConstraint}
	y \simeq 3 g_*^{1/4} \frac{\Treh}{\sqrt{m_\varphi M_{pl}}}.
\end{eqnarray}
This choice of coupling corresponds to approach \ref{it:approachA}, i.e., avoiding backreaction. In the following we apply \eqref{YukawaConstraint} without further restrictions to the models introduced in Sec.~\ref{sec:ColdModel} .

\subsection{Warm inflation}\label{sec:warm}
\emph{Warm inflation} \cite{Berera:1995ie} is an alternative incarnation of the inflationary paradigm in which dissipation is efficient enough to maintain a thermal bath with temperature $T>H$ during inflation, cf.~\cite{Kamali:2023lzq} for a review.
It corresponds to a quasi-stationary solution of the coupled set of equations \eqref{EOM} for which the Hubble dilution of the radiation density $\rho_R$ is exactly compensated by particle production through inflaton dissipation. Assuming fast thermalisation of the produced particles, this implies the presence of a thermal bath of temperature 
\begin{eqnarray}\label{Twarm}
    \frac{\pi^2}{30} g_{*} T^4  \approx \frac{\GG}{4H} \left(\frac{\V'}{3H + \GG}\right)^2. 
\end{eqnarray}

Conceptually warm inflation differs from standard cold inflation in two ways. Firstly, the friction term in \eqref{EOM} affects the slow roll dynamics, modifying the slow roll parameters to 
$\epsilon = \left(\V'/\V \right)^2 M^2_{\text{pl}}/[2(1 +Q)] $ , $\eta = (\V''/\V) M^2_{\text{pl}}/(1 +Q) $
with $Q = \GG/(3H)$. Secondly, thermal fluctuations contribute to sourcing scalar cosmological perturbations. As a result, for fixed $\V$ warm inflation predicts a smaller $r$. 
It is common to distinguish two regimes of warm inflation. In the strong regime with $Q>1$ the thermal bath affects both the evolution of the background field $\varphi$ and the generation of cosmological perturbations \footnote{In cold inflation $\GG < H$ during both inflation and reheating, and reheating ends when $\GG$ exceeds $H$.
During warm inflation $\GG>H$ does not imply an imminent transition to radiation domination. 
The reason is that $\GG > H$ leads to an efficient depletion of $\rho_\varphi$ into $\rho_R$ when $\varphi$ performs oscillatory motion, but this is not necessarily true during slow roll.}; in the weak regime with $Q<1$ its impact on the background field evolution is subdominant, but it still sources cosmological perturbations.

A major challenge when implementing this idea within the framework of realistic quantum field theories lies in constructing a model in which inflaton interactions are strong enough to generate a sizable dissipation term $\GG$ while at the same time avoiding thermal corrections to $\V$ that would spoil the slow roll dynamics \cite{Yokoyama:1998ju}.  
For the present analysis we solve this problem by assuming an axion-like inflaton coupling to gluons \cite{Berghaus:2025dqi},
for which sphaleron heating \cite{McLerran:1990de} yields a friction term of the form \footnote{The suppression of the friction term by the chiral chemical potentials of light fermions is alleviated by their Hubble dilution \cite{Drewes:2023khq,Berghaus:2025dqi}.}
\begin{eqnarray}\label{DissipationWarm}
\GG =  N_c^5 \upalpha_s^5 \frac{T^3}{2f^2} \Big/\left(1+\frac{2 N_f N_c^4 \alpha^5_s T}{ \sqrt{\V/(3M^2_{\text{pl}}})}\right)
\end{eqnarray}
while the axion's 
\footnote{We adapt the term \emph{axion} here to emphasize that the field $\varphi$ in \eqref{eq:lag} behaves like a standard QCD axion from a phenomenological viewpoint, in spite of the fact that the addition of the potential $\V$ alters its fundamental properties \cite{Zell:2024cyz}.}
shift symmetry protects $\V$ from thermal corrections \cite{Berghaus:2019whh}.
Here $\upalpha_s = g_s^2/(4\pi)$ with $g_s$ the strong gauge coupling constant in the SM and $f$ is the axion decay constant. 
$N_c=3$ and $N_f=5$ the effective numbers of colors and flavors in the SM, respectively. 
Hence, the inflaton's interactions with other fields are given by one single parameter $f$
and since this coupling mediates an interaction with gluons, it is directly accessible in axion search experiments.

\section{Theoretical Benchmark Models}\label{sec:model}

We assess AliCPT-1's sensitivity to reheating in several models of inflation.

\subsection{Standard single field inflation}\label{sec:ColdModel}
We first consider standard (cold) inflation, cf.~Sec.~\ref{sec:cold}, 
in two benchmark scenarios.
Both are so-called plateau models with two parameters $\{\M,\alpha\}$ in the potential $\V$.
Here $\M$ is the scale of inflation and $\alpha$ a dimensionless parameter that determines the strengths of all inflaton self-interactions 
as well as the ratio between $\M$ and the inflaton mass $m_\varphi$. 
To avoid a dependence of the CMB spectra on a number of fundamental parameters that exceeds the number of observables $\{A_s, n_s, r\}$, cf.~challenges \ref{challenge1} and \ref{challenge2}, 
we pursue the approach \ref{it:approachA} of avoiding backreaction by focusing on scenarios in which reheating is driven by a Yukawa interaction. In this case $\Treh$ can be linked to a single  microphysical parameter $y$ by \eqref{YukawaConstraint}.

\subsubsection{
$\alpha$-attractor T-model
}\label{sec:aT}
\begin{figure}
	\centering
	    \includegraphics[width=0.9\linewidth]{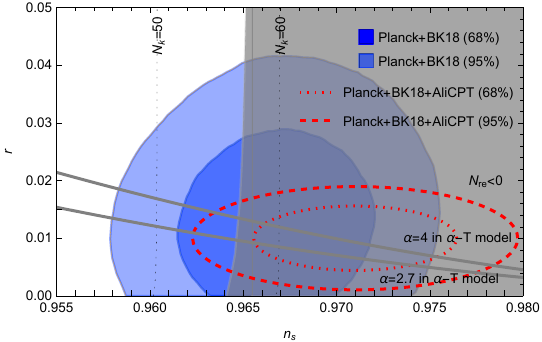}
    \includegraphics[width=0.9\linewidth]{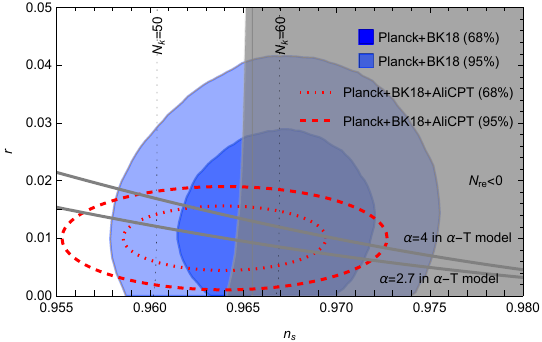}
	\caption{The predictions of the $\alpha$-T model \eqref{alpha V} for $n_s$ and $r$
        lie on the lines defined by \eqref{AlphaInAlphaT}. The position along this line is determined by $\Nreh$, hence each point corresponds to a value of $\Treh$ obtained from \eqref{Tre}. The gray region on the right  corresponds to $N_{\rm re} < 0$. The dotted lines correspond to $N_k=50$ and $N_k=60$
        $e$-folds between the horizon-crossing of the pivot scale $k=0.05 \ {\rm Mpc}^{-1}$ and the end of inflation. The blue areas display  marginalized joint confidence regions for $(n_s, r)$ at the $1\sigma$ $(68\%)$ CL (dark) and $2\sigma$ $(95\%)$ CL (light) from Planck+BICEP/Keck 2018 data (Fig.~5 in~\cite{BICEP:2021xfz}). The red lines indicate the $1\sigma$ and $2\sigma$ contours of the likelihood function \eqref{Eq:Likelihood} in benchmarks \ref{it:A} (upper panel) and \ref{it:B} (lower panel) in table \ref{PrincipalBenchmark}.
        } 
	\label{fig:ellipsesalphaT}
\end{figure}

The popular $\alpha$-attractor T-model ($\alpha$-T)~\cite{Kallosh:2013maa,Kallosh:2013hoa,Carrasco:2015pla,Carrasco:2015rva} 
is described by the potential
\begin{eqnarray}
	\V&=&\M^4{\rm tanh}^{2
	}
	\left(\frac{\varphi}{\sqrt{6\alpha}M_{pl}}\right).
	\label{alpha V} 
\end{eqnarray}
Solving $\epsilon=1$ for $\varphi$ gives
\begin{eqnarray}
	\varphi_{\rm end} = \frac{M_{pl}}{2}\sqrt{\frac{3\alpha}{2}}{\rm ln}\left(
	\frac{3\alpha + 8 + 4\sqrt{4^2 + 3\alpha}}{3\alpha}
	\right). 
\end{eqnarray}
The scale of inflation $\M$ is obtained from \eqref{H_k} as
\begin{equation}
	\M =
	M_{pl}
	\left(
	\frac{3\pi^2}{2}
	A_s r
	\right)^{1/4}
	\tanh^{-\frac{\nn}{2}}\left(
	\frac{\varphi_k}{\sqrt{6\alpha}M_{pl}}
	\right), \quad \
	\label{M3}
\end{equation}
while
\begin{equation}
	m_\varphi=\frac{M^2}{\sqrt{3\alpha}M_{pl}} \ , \
	g_\phi = 0 \ , \
	\lambdaphi =-\frac{4M^4}{9\alpha^2M^4_{pl}},
	\label{alphaTlambda}
\end{equation}
and
\eqref{TakaTukaUltras} 
yields the relation
\begin{equation}\label{AlphaInAlphaT}
	\alpha=\frac{4r}{3(1-n_s)(4(1-n_s)-r)},
\end{equation}
which defines a line in the plane spanned by $n_s$ and $r$, see Fig.~\ref{fig:ellipsesalphaT}. This illustrates that fixing the inflationary model \eqref{alpha V} is insufficient to uniquely predict $n_s$ and $r$, as the position along the line defined by \eqref{AlphaInAlphaT} is determined by the duration of the reheating epoch $\Nreh$. 

\begin{figure}
	\centering
    \includegraphics[width=0.9\linewidth]{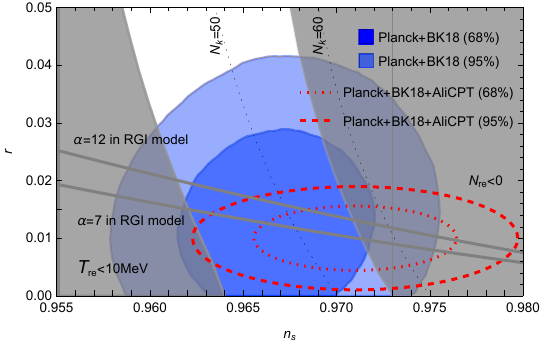}
	\caption{Predictions from the RGI model \ref{RGI V}, with gray lines defined by \eqref{Nepenthe} and all other conventions analogue to Fig.~\ref{fig:ellipsesalphaT}.
    } 
	\label{fig:ellipsesRGI}
\end{figure}

\begin{figure}
	\centering
	\includegraphics[width=0.9\linewidth]{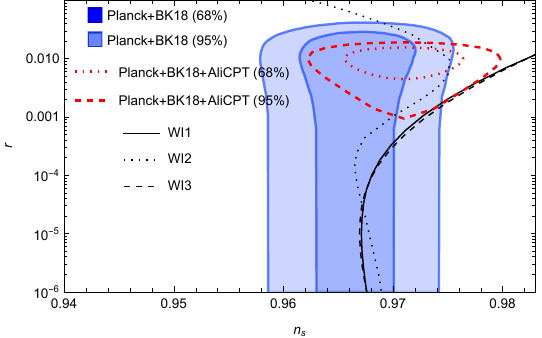}
    \includegraphics[width=0.9\linewidth]{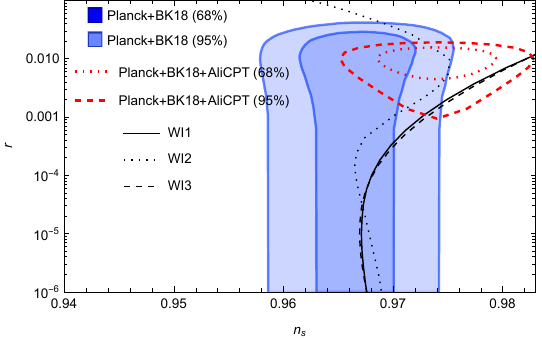}
	\caption{After imposing the current measurement
 of $A_s$ the warm inflation model \eqref{eq:lag} contains only one free combination of parameters, which defines a line in the $n_s$-$r$ plane. 
However, the precise shape of this line and the relation to the model parameters still suffer from theoretical uncertainties that are comparable in to the current observational uncertainty in $n_s$. 
To illustrate this, we plot the results of three different computational methods labeled as WI1, WI2 and WI3, cf.~Table~\ref{WIcomputations}.
 We compare this to current Planck+BK constraints \cite{BICEP:2021xfz} (in blue) and the expected sensitivity from AliCPT-1  in Table \ref{PrincipalBenchmark} (red ellipses corresponding to benchmark \ref{it:A} in the upper panel and \ref{it:C} in the lower panel). 
    }
	\label{fig:warmellipses}
\end{figure}

\subsubsection{
RGI-model
}\label{sec:RGI}

The radion gauge inflation (RGI)~\cite{Fairbairn:2003yx,Martin:2013nzq} model is defined by the potential
\begin{eqnarray}\label{RGI V}
    \V = \M^4\frac{(\varphi/M_{pl})^2}{\alpha+(\varphi/M_{pl})^2}, 
\end{eqnarray}
which yields
\begin{equation}\label{InflatonMassRGI}
    m_\varphi=\sqrt{\frac{2}{\alpha}}\frac{M^2}{M_{pl}}~,\quad 
    g_\phi=0~,\quad
    \lambdaphi =-\frac{24M^4}{\alpha^2M^4_{pl}}~.
\end{equation}
The scale of inflation $\M$  from \eqref{H_k} is
\begin{equation}
    M =
     M_{pl}
     \left(
     \frac{3\pi^2}{2} r A_s
     \left(1 + \alpha \frac{M^2_{pl}}{\varphi_k^2}\right)
     \right)^{1/4}.
    \label{M2}
\end{equation}
As for the $\alpha$-T model, fixing $\alpha$ defines a line in the $n_s$-$r$ plane, cf.~Fig.~\ref{fig:ellipsesRGI},
\begin{equation}\label{Nepenthe}
    \alpha=\frac{432r^2}{(8(1-n_s)+r)^2(4(1-n_s)-r)}.
\end{equation}
The end of inflation at leading order in the slow roll parameters is given by
\begin{eqnarray}
\varphi_{\rm end} = M_{pl}
\frac{\xx^{1/3} - \alpha}{\sqrt{3} \xx^{1/6}}
\label{phiendRGI}
\end{eqnarray}
with $\xx=\alpha^2(27+\alpha+3\sqrt{81+6\alpha})$.

\subsection{QCD-driven warm inflation:
$\varphi^4$-model}\label{sec:WarmModel}

In a second example we pursue the approach \ref{it:approachB} to address both challenges \ref{challenge1} and \ref{challenge2}: 
In order to tackle the problem that the number of unknown parameters (those in $\V$ and those related to the reheating process) exceeds the number of observables $\{A_s, n_s, r\}$, we focus on a minimal model with only one parameter $\lambdaphi $ in the potential 
\footnote{Note that the normalization of the coefficient $\lambda$ in \eqref{eq:lag} differs from that of the standard Taylor expansion used in \eqref{alphaTlambda} and \eqref{InflatonMassRGI} by a factor $4!$ in order to be consistent with the literature based on which we computed the CMB perturbations, see Table \ref{WIcomputations} below. }
and a single  inflaton coupling 
$\g=\upalpha_s m_\varphi/(8\pi f)$
to the gluon field strength tensor \cite{Berghaus:2025dqi}, 
\begin{equation}
\label{eq:lag}
\mathcal{L} = \mathcal{L}_{\rm SM} + \frac{1}{2} \partial^\mu \varphi \partial_\mu \varphi - \lambda \varphi^4  
- \frac{\upalpha_s}{8 \pi}\frac{\varphi}{f} G^{a}_{\mu \nu} \tilde{G}^{a\mu \nu},
\end{equation}
with $\mathcal{L}_{\rm SM}$ the SM Lagrangian. 
While this type of potential is ruled out observationally in cold inflation (where quantum fluctuations are the sole source of cosmic perturbations), it is consistent with present data if the thermal fluctuations seen in warm inflation scenarios are added \cite{Mirbabayi:2022cbt,Berghaus:2025dqi}.

The Lagrangian \eqref{eq:lag} describes a complete model of inflation and reheating.
It is minimal in two senses. 
Firstly, it can reproduce the current CMB observations with a simple monomial potential with a single free parameter.
Secondly, it does not require any mediator particles with unknown properties to fuel reheating, as the inflaton directly couples to SM fields.
After fitting the observed value of $A_s$, the model contains only one unconstrained parameter, a combination of $f$ and $\lambda$ that has to be determined numerically, which defines a line in the $n_s$-$r$ plane shown in Fig.~\ref{fig:warmellipses}.
This sole free parameter can be measured by an observation of $r$. Moreover, the inflaton itself can be searched for in axion experiments and its coupling can be measured, 
offering a unique opportunity to probe the connection between big bang cosmology and particle physics by combining future CMB observations with laboratory searches.

The expansion history of the universe can in principle be fully reconstructed in the model \eqref{eq:lag}, but in practice this requires knowledge of $\GG$ along the entire trajectory of $\varphi$. 
Established computations are only available during slow roll and oscillations near the minimum of $\V$, see e.g.~\cite{Laine:2021ego,DEramo:2021lgb,Laine:2025rll} and references therein. 
While the continuous presence of a thermal bath in warm inflation scenarios implies that there is no separate reheating period in the conventional sense, the moments when inflation ends ($\ddot{a} < 0$) and when radiation dominates ($\rho_R > \dot{\varphi}/2 + \V$) could be separated by a number of $e$-folds $\Nreh$ that leaves an imprint in the CMB. However, a simple estimate suggests that this period would be unobservably short even if \eqref{DissipationWarm} held at all times. Hence, we can in first approximation identify $\Treh$ with \eqref{Twarm}. For warm inflation we cannot use \eqref{CouplinfToTreh} to make connection to the microphysical coupling constant, but instead rely on the numerical relations between the parameters obtained from the approaches listed Table \ref{WIcomputations}.

\section{ Sensitivity Forecasts }\label{sec:forecasts}

The sensitivity of CMB observatories to the reheating temperature $\Treh$ and the inflaton coupling $\g$ can be related to their sensitivity to the scalar spectral index $n_s$ and the scalar-to-tensor ratio $r$. 
In the case of AliCPT-1, the main improvement compared to combined Planck and BICEP/Keck (Planck+BK) data 
\cite{BICEP:2021xfz} comes from the improvement in the error bar $\sigma_r$ that can enable a discovery of primordial B-modes.

\subsection{Combining AliCPT-1  and Planck}\label{sec:sims}

We follow the approach outlined in Ref.~\cite{Errard:2015cxa}, 
where a comprehensive description of the methodology is provided. Given the experimental configurations detailed in Table \ref{alicpt-config}, we first perform diffuse foreground rejection to obtain the residual foreground power spectra. Next, we apply delensing processing to derive the delensed $BB$ spectra. Finally, we use the Fisher matrix to estimate the scientific parameters. 
Our analysis includes the standard six $\Lambda$CDM parameters: $\{n_s, A_s, \tau, H_0, \Omega_bh^2,  \Omega_ch^2\}$. Additionally, we consider extended models, including the ratio of the primordial tensor to scalar power spectra, $r$. We set the fiducial values standard six $\Lambda$CDM parameters to the best-fit $\Lambda$CDM values derived from Planck’s analysis \cite{Planck:2015fie} and $r$ to $0.01$.
We take the configuration of Planck from \cite{Errard:2015cxa}.

\begin{table}
	\centering
	\begin{tabular}{c||c|c|c|c}
		Frequency [GHz] & 27 & 40 & 90 & 150 \\
		\hline
		FWHM [arcmin] & 97.07 & 65.5 & 16.2 & 9.7 \\ 
		\hline
		map-depth [$\mu K$arcmin] & 50 & 35 & 2.2 & 3.3 \\
		\hline
		$\alpha_{knee}$ & -2.4 & -2.4 & -2.5 & -3.0 \\
		\hline
		$\ell_{knee}$ & 30 & 30 & 50 &  50 \\
	\end{tabular} 
	\caption{Configurations for AliCPT-1, the $\alpha_{knee}$ and $\ell_{knee}$ are used to describe the $1/f$ noise, following $N_l=N_{white}(1+(\ell/\ell_{knee})^{\alpha_{knee}})$.}
	\label{alicpt-config}
\end{table}

In the Fisher matrix framework, the likelihood is assumed Gaussian, and marginalization over other parameters yields a $2\times2$ covariance matrix for $r$ and $n_s$. The off-diagonal elements are more than an order of magnitude smaller than the diagonal ones, and we have verified that setting them to zero has negligible impact on our final results. We therefore adopt a diagonal approximation in subsequent calculations.
In terms of one-dimensional marginalized standard deviations for $n_s$ and $r$ we obtain
$\sigmans=0.0032$ and $\sigmar=0.0037$ for the fiducial values $\nsbar=0.967$ and $\rbar=0.01$.

Our analysis is based on a set of idealized assumptions that facilitate a clear assessment of statistical sensitivity. 
For example, we model the instrumental noise as homogeneous across the sky for both \textit{Planck} and AliCPT, and for AliCPT we account for $1/f$ noise only at the level of the angular power spectrum. 
Furthermore, the forecast is derived within the Fisher matrix framework, rather than through full end-to-end simulations and Markov Chain Monte Carlo inference.

While AliCPT will deliver high-sensitivity E- and B-mode polarization data over its survey region, 
its impact on the constraint of the scalar spectral index $n_s$ is expected to be limited. 
Including $1/f$ noise, its sensitivity to E-mode polarization is comparable to that of current 
ground-based experiments, which---combined with \textit{Planck}---already provide strong constraints on $n_s$. 
Accordingly, our Fisher forecast shows only a marginal improvement in $\sigma_{n_s}$ when adding AliCPT to 
\textit{Planck}. 
In contrast, AliCPT’s primary scientific value lies in its enhanced sensitivity to primordial B-modes, 
offering a significant step forward in constraining the tensor-to-scalar ratio $r$. In light of this modest expected improvement on $\sigma_{n_s}$ and to maintain consistency with the latest 
observational results, we adopt the value of $\sigma_{n_s}$ reported in Fig.~5 of~\cite{BICEP:2021xfz} 
for our analysis.

Hence, for the purpose of our analysis, we effectively characterize the expected information gain from adding AliCPT-1  data to existing constraints from Planck 
by a two-dimensional Gaussian $\mathcal{N}(n_s,r|\nsbar,\sigmans;\rbar,\sigmar)$ in the $n_s$-$r$ plane.
Our principal observational benchmark is defined by the central values and variances labeled as \ref{it:A} in Table \ref{PrincipalBenchmark},  the corresponding 68\% and 95\% confidence regions are shown in  Fig~\ref{fig:ns_vs_r}.
Our primary goal is to investigate how much such an observation could reveal about $\Treh$ and the inflaton couplings in the models \eqref{alpha V}, \eqref{RGI V} and \eqref{eq:lag}.

The observational benchmarks \ref{it:B} and \ref{it:C} correspond to fiducial values $\nsbar$ that are more in line with the predictions of the models \eqref{alpha V} and \eqref{eq:lag}, respectively, and which we study in addition for those models. 
Note that the error bar of $n_s$ does not strongly depend on its fiducial value within the range of values for $\nsbar$ considered here, justifying to use the same $\sigmans$ for all three fiducial values \ref{it:A}--\ref{it:C}. The same would not be true for $\sigmar$, which strongly depends on $\rbar$. Roughly speaking, the results of our simulation can be extrapolated to define new observational benchmarks by shifting the ellipse in Fig.~\ref{fig:ns_vs_r} horizontally, but shifting it vertically is not allowed.

We refer to the sensitivities presented in table \ref{PrincipalBenchmark} as AliCPT+Planck in the following.

\begin{table}
\centering
\begin{tabular}{c | c | c | c | c | c}
\begin{tabular}{c}
benchmark
\end{tabular} & $A_s$ & $\nsbar$ & $\sigmans$ & $\rbar$ & $\sigmar$\\
\hline 
Planck+BK & $10^{-10}e^{3.043}$ &  0.967  & 0.0036 & (0.01) &  (0.012) \\
\begin{enumerate*}[A]
\item \label{it:A}
\end{enumerate*}
& $10^{-10}e^{3.043}$ &  0.971  & 0.0036 & 0.01 & 0.0037  \\
\begin{enumerate*}[B]
\item \label{it:B}
\end{enumerate*} & $10^{-10}e^{3.043}$ & 0.964  & 0.0036 & 0.01 & 0.0037 \\
\begin{enumerate*}[C]
\item \label{it:C}
\end{enumerate*} & $10^{-10}e^{3.043}$ & 0.974  & 0.0036 & 0.01 & 0.0037 
\end{tabular}
\caption{Fiducial values and uncertainties that define our observational benchmark scenarios, compared to the parameters fitted to the Planck+BK results, obtained from Fig.~5 in \cite{BICEP:2021xfz}. The error on $A_s$ can be neglected compared to the other uncertainties \cite{Drewes:2023bbs}, hence we fix $A_s$ to its best fit value from \cite{Planck:2018vyg}.
}
\label{PrincipalBenchmark}
\end{table}

\subsection{Statistical approach}

With three observables $\{A_s, n_s, r\}$ at hand we can in principle measure up to three parameters, for which we may e.g.~choose the three physical scales $\{\M,m_\varphi,\Treh\}$. The perspectives to perform these measurements in practice are, however, model dependent.

\begin{itemize}
\item The cold benchmark models \eqref{alpha V} and \eqref{RGI V} contain two parameters $\{\M,\alpha\}$ in the potential $\V$. Hence, at the microphysical level there are three dimensionless free parameters $\{\M/M_{pl},\alpha,y\}$ that determine the three scales $\{\M,m_\varphi,\Treh\}$. 
After fixing $A_s$, cf. Table \ref{PrincipalBenchmark}, the scale of inflation $\M$ is essentially determined by observing $r$, see \eqref{M3} and \eqref{M2}. 
However, Figs.~\ref{fig:ellipsesalphaT} and \ref{fig:ellipsesRGI} show that $\sigmans$ is too large to measure the dimensionless parameter $\alpha$ \footnote{A detailed discussion of this point can be found in ~\cite{Drewes:2023bbs}.}.
We address this challenge \ref{challenge1} with two approaches. In the first we define theoretical benchmark scenarios with fixed $\alpha$, in the second we express the knowledge gain in terms of a two-dimensional posterior in the $\alpha$-$\Treh$ plane.
In the former case we fix $\alpha$ to  values that are consistent with the condition \eqref{GeneralScalingSelf} and use $\x\equiv \log_{10}y$ as the only remaining free parameter that is to be constrained from data. In the latter case we use $\x$ and $\alpha$ as free parameters.

\item The warm inflation model \eqref{eq:lag} contains only two free parameters $\lambda$ and $f$ which together determine the dynamics all the way from inflation to radiation domination, with no clear separation between inflation and reheating. Hence, the number of observables $\{A_s, n_s, r\}$ exceeds the number of model parameters, and the size of $\sigmans$ does not necessitate any additional model assumptions in this scenario. 
This minimality permits to avoid the two challenges \ref{challenge1} and \ref{challenge2} related to the number of parameters in $\V$ and during reheating at once.
Fixing $A_s$ leaves us with one unconstrained combination $\x$ of parameters, 
which defines a line in the $n_s$-$r$ plane shown in Fig.~\ref{fig:warmellipses}.
The relation between $\lambda$, $f$ and $\x$ has to be determined numerically, cf.~Table \ref{WIcomputations}.
\end{itemize}
Hence, depending on which of the above cases we consider, the set of free model parameters $\X$ to be constrained from data either consists only of one number $\X=\x$ or of the pair $\X=\{\x,\alpha\}$.

In the following we employ the simple analytic method introduced in \cite{Drewes:2022nhu} to make connection between the unknown parameters and observables, which was shown to be conservative in \cite{Drewes:2023bbs}.  
Prior to CMB observations it is known that
$\Nreh > 0$ and that there is a lower bound $\Treh > T_{\rm BBN}$ to allow for successful primordial nucleosynthesis \cite{Cyburt:2015mya},
motivating to the prior probability density function 
\begin{eqnarray}\label{PxdefNew}
	P(\X)  
	= C_1 \theta\big(\Treh(\X)-T_{\rm BBN}\big)
	\ \theta\big(\Nreh(\X)\big),
\end{eqnarray}
where $\theta$ denotes the Heaviside function,  
and the constant $C_1$ is determined by the requirement $\int d\X P(\X) = 1$.
In the present work we use $T_{\rm BBN} = 10$ MeV.

We can quantify the knowledge gain about $\X$  obtained from data $\mathcal{D} = \{n_s,r\}$ in terms of a posterior distribution $P(\X|\mathcal{D})=P(\mathcal{D}|\X)P(\X)/P(\mathcal{D})$, where $P(\mathcal{D})=\int d\X P(\mathcal{D}|\X)P(\X)$,
\begin{eqnarray}\label{Eq:Likelihood}
	P(\mathcal{D}|\X) = C_2\mathcal{N}(n_s,r|\nsbar,\sigmans;\rbar,\sigmar)\theta(r). 
\end{eqnarray}
The constant $C_2$ is fixed by $\int P(\mathcal{D}|\X)d\mathcal{D} = 1$. 
The function $\mathcal{N}(n_s,r|\nsbar,\sigmans;\rbar,\sigmar)$ contains the information about the experimental sensitivities $\sigmans$ and $\sigmar$ for given fiducial values $r=\rbar$ and $n_s=\nsbar$.

\subsection{Results}

In the following we quantify the information gain on reheating and the related fundamental physics that can be expected from adding AliCPT-1  data to existing Planck observation.

\subsubsection{Standard cold inflation}
In the two cold inflation models $\alpha$-T and RGI, \eqref{alpha V} and \eqref{RGI V}, 
the scale of inflation $\M$ can be constrained by a measurement of $r$. However, the current error $\sigmans$ is too small to pin down the parameter $\alpha$. AliCPT-1  will not fundamentally change this situation, but can nevertheless provide important information gain on combinations of $\alpha$ and  $\Treh$. For the example of the Yukawa coupling $y$ considered here, the latter can be converted into knowledge on $\x=\log_{10}y$.

\begin{table}
	\centering
    \begin{tabular}{c|c|c|c|c|c}
		 model & $\alpha$ & benchmark & $\x$ & 
           $10^3 \M/M_{pl}$
         & ${\rm log}_{10}(\frac{T_{\text{re}}}{\text{GeV}})$\\
		\hline
		RGI & 7 & \ref{it:A} & $-4.9\pm3.7$ & $4.46\pm0.10$ & $10.2\pm3.7$ \\
		 RGI & 12 & \ref{it:A} &  $-3.4\pm3.0$ & $4.81\pm0.09$ & $11.6\pm3.0$ \\
      $\alpha$-T & 2.7 & \ref{it:A} &  $-2.2\pm2.2$ & $4.31\pm0.07$ & $12.8\pm2.2$ \\
      $\alpha$-T & 4 & \ref{it:A} & $-1.7\pm1.8$ & $4.71\pm0.06$ & $13.3\pm1.8$ \\
            $\alpha$-T & 2.7 & \ref{it:B} &  $-4.4\pm3.3$ & $4.38\pm0.11$ & $10.6\pm3.3$ \\
      $\alpha$-T & 4 & \ref{it:B} & $-3.1\pm2.6$ & $4.75\pm0.09$ & $11.9\pm2.7$ 
	\end{tabular}
	\caption{One-dimensional posterior mean values and variances for $\M$, $\x$ and $\Treh$ in the two theoretical models \eqref{alpha V} and \eqref{RGI V} for the observational benchmarks defined in Table \ref{PrincipalBenchmark}.
        The intervals corresponding 68\% of the probability around the medians are ${\rm log}_{10}(T_{\text{re}}[\text{GeV}]) \in [6.2, 14.0] $,  $ [8.5, 14.6] $, $[10.7, 14.9]$, $[11.6, 15.0]$, $[7.1, 14.1]$ and $[9.2,14.6]$. }
	\label{ErrorBars}
\end{table}

\begin{figure}
	\centering
	\includegraphics[width=0.9\linewidth]{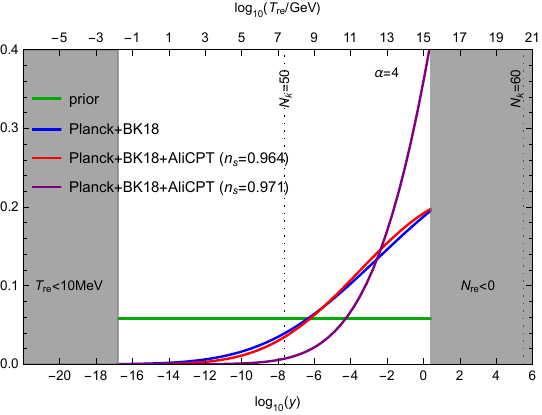}
	\includegraphics[width=0.9\linewidth]{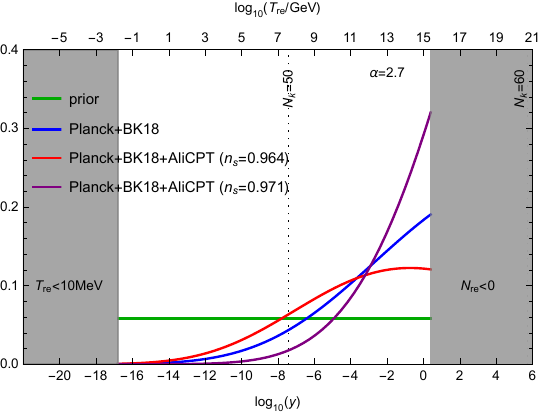}
	\caption{Posterior distributions for the reheating temperature $\Treh$ and Yukawa coupling $y$
    in the $\alpha$-attractor model \eqref{alpha V} with fixed $\alpha=4$ (upper panel) and $\alpha=2.7$ (lower panel) 
    for the benchmarks \ref{it:A} and \ref{it:B}. 
    The dark gray area on the left is excluded by BBN, that on the right by the requirement that energy density during reheating cannot exceed the scale of inflation.
		The standard deviations for the posteriors are summarized in Table \ref{ErrorBars}.} 
	\label{fig:posteriorsaT}
\end{figure}

We first consider the $\alpha$-attractor model \eqref{alpha V}. 
Fig.~\ref{fig:posteriorsaT} shows one-dimensional posteriors in $\x$ for fixed $\alpha$, 
the corresponding error bars on $\M$, $\x$ and $\Treh$ are given in Table \ref{ErrorBars}. 
The values of $\alpha$ were chosen such that the line defined by \eqref{AlphaInAlphaT} crosses the Planck+BK 1$\sigma$ region in Fig.~\ref{fig:ellipsesalphaT} while being consistent with the condition \eqref{GeneralScalingSelf}.
For our principal observational benchmark \ref{it:A} defined in Table \ref{PrincipalBenchmark}, 
the posterior variance for $\alpha=4$ improves from  $\sigmax =  2.9$ to  $\sigmax =  1.8$ when comparing to Planck+BK, while for $\alpha=2.7$ it  improves from $\sigmax =  3.0$ to  $\sigmax =  2.2$.
This improvement is largely due to the fact that the observational benchmark \ref{it:A} would almost rule out the model \eqref{alpha V}  at the $2\sigma$ level, as most of the observationally preferred parameter choices would fall into the forbidden $\Nreh<0$ region. 
This can be seen in Fig.~\ref{fig:ellipsesalphaT}; it is also reflected by the Bayes factors 0.16 and 0.39, respectively, when comparing benchmark \ref{it:A} to Planck+BK values. 
For the observational benchmark \ref{it:B}, the model can provide a much better fit to the data, but there is little or no improvement in $\sigmax$. 

\begin{figure}
	\centering
    \includegraphics[width=0.9\linewidth]{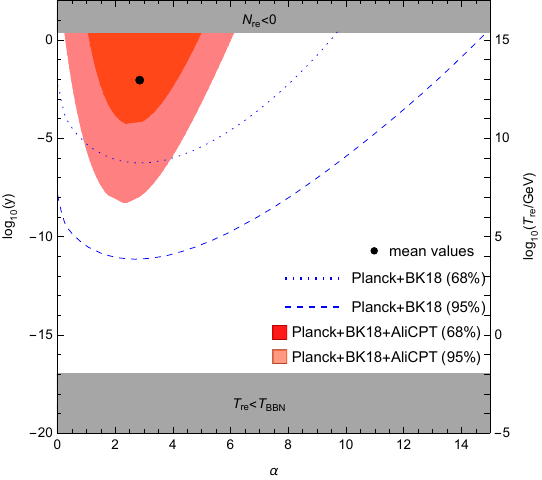}
	\includegraphics[width=0.9\linewidth]{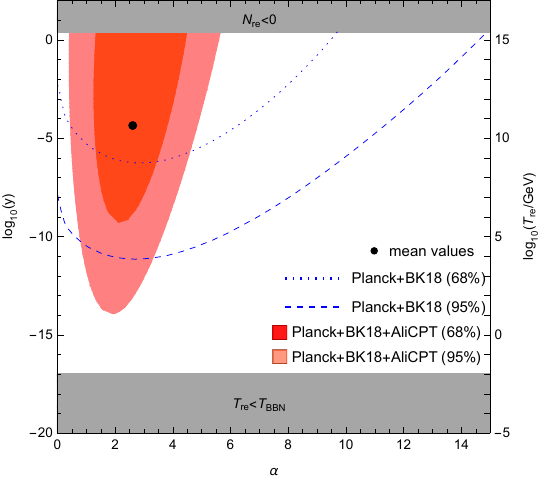}
	\caption{Two-dimensional posteriors in $\alpha$ and $\Treh$ with corresponding $y$ 
    for the $\alpha$-T model \eqref{alpha V}
    in benchmarks \ref{it:A} (upper panel) and \ref{it:B} (lower panel) compared to current constraints from Planck+BK. The gray areas indicate the regions where $\Nreh<0$ and $T < T_{\rm BBN}$. } 
	\label{fig:2DpostalphaT}
\end{figure}

Since $\alpha$ cannot be predicted from theory, it is instructive to quantify the knowledge gain that can be achieved with AliCPT-1  in terms of the two-dimensional posteriors displayed in Fig.~\ref{fig:2DpostalphaT}.
Evidently AliCPT-1  can considerably narrow the posterior for both observational benchmarks \ref{it:A} and \ref{it:B} but the improvement is better in case \ref{it:A}. 
In this case the line defined by \eqref{AlphaInAlphaT} crosses the observationally preferred region in Fig.~\ref{fig:ellipsesalphaT} only for a comparably narrow interval of values $1\lesssim \alpha \lesssim 5$, and within this region  $\Treh >  10^{10}$ GeV is preferred.
In summary, in the $\alpha$-attractor model \eqref{alpha V} AliCPT-1's potential to gain information on reheating by measuring $r$ strongly depends on the value of $n_s$.
For both observational benchmarks \ref{it:A} and \ref{it:B} AliCPT-1  can considerably narrow down the range of allowed values for $\alpha$. 
In case \ref{it:A} AliCPT-1  could tighten the current lower bound on $\Treh$ from Planck+BK by two orders of magnitude.
At the $1\sigma$ level, $\alpha$ is, roughly speaking, constrained to the range $1\lesssim \alpha \lesssim 5$.
If reheating proceeds through a Yukawa interaction, the coupling constant $y$  must be larger than the electron Yukawa coupling in the SM for $\alpha\simeq 2$, and of order unity at the upper and lower ends of the interval $1\lesssim \alpha \lesssim 5$.

\begin{figure}
	\centering
	\includegraphics[width=0.9\linewidth]{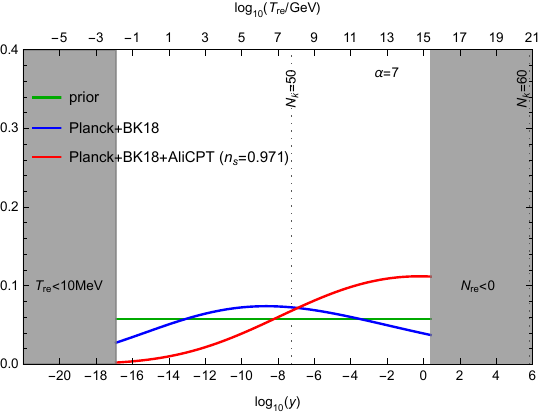}
	\includegraphics[width=0.9\linewidth]{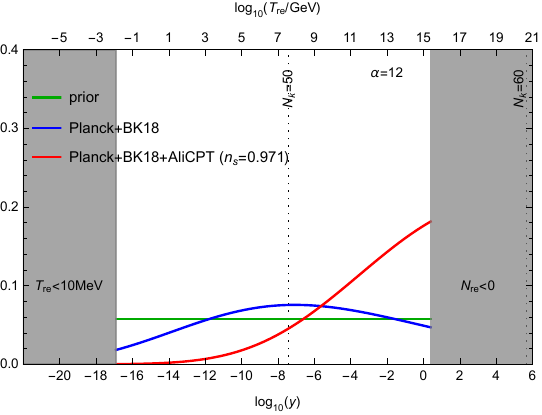}
	\caption{Posterior distributions for the reheating temperature $\Treh$ and Yukawa coupling $y$
    in the RGI model \eqref{RGI V} with fixed $\alpha=7$ (upper panel) and $\alpha=12$ (lower panel) 
    for the benchmark \ref{it:A}, with all other conventions as in Fig.~\ref{fig:posteriorsaT}.
    } 
	\label{fig:posteriorsRGI}
\end{figure}

\begin{figure}
	\centering
	\includegraphics[width=0.9\linewidth]{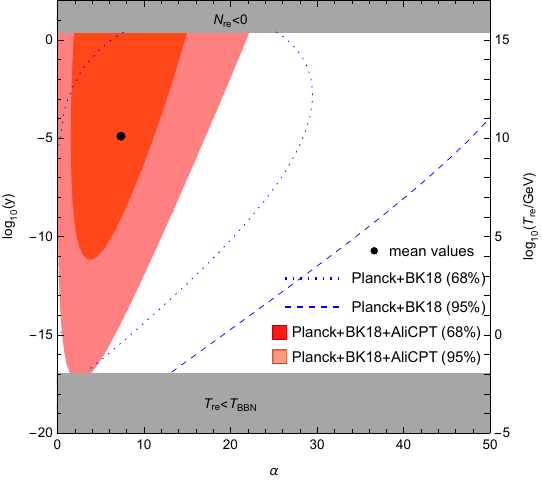}
	\caption{Two-dimensional posteriors in $\alpha$ and $\Treh$ with corresponding $y$ 
    for the RGI model \eqref{RGI V}
    in benchmark \ref{it:A} compared to Planck+BK, with conventions as in Fig.~\ref{fig:2DpostalphaT}.
    } 
	\label{fig:2DpostRGI}
\end{figure}

Turning to the RGI model \eqref{RGI V}, 
it is already evident from Fig.~\ref{fig:ellipsesRGI} that it can provide a better fit for our observational benchmark \ref{it:A}. On the other hand, this also implies less knowledge gain on reheating, simply because model is not as cornered as the $\alpha$-attractor model \eqref{alpha V}.
Nevertheless, depending on the value of $\alpha$, AliCPT-1 could reduce the  $\sigma_\x = 4.4$ from Planck+BK to  $\sigma_\x = 3.0$, 
cf.~Table \ref{ErrorBars}. 
The corresponding one-dimensional posteriors are shown in Fig.~\ref{fig:posteriorsRGI}.
Fig.~\ref{fig:2DpostRGI} displays the two-dimensional posterior, which shows that AliCPT-1's potential to constrain reheating strongly depends on the value of $\alpha$. This can easily be understood from Fig.~\ref{fig:ellipsesRGI}: For larger $\alpha$, the fraction of the line defined by \eqref{Nepenthe} that falls into the observationally preferred region decreases.

\subsubsection{QCD-driven warm inflation}

\begin{table}
	\centering
	\begin{tabular}{c||c|c|c}
		method& WI1 & WI2 & WI3  \\
		\hline
		reference & \cite{Berghaus:2025dqi}, & \cite{ORamos:2025uqs} & \cite{Laine:2025rll} 
	\end{tabular} 
	\caption{In the warm inflation model \eqref{eq:lag} the relations between the observables $\{A_s, n_s, r\}$ and the model parameters $\{f,\lambda\}$ can only be determined numerically. 
The computation of the spectrum of cosmic perturbations for warm inflation is still an area of active research \cite{Mirbabayi:2022cbt, Montefalcone:2023pvh,Berghaus:2024zfg,Laine:2024wyv,Rodrigues:2025neh,ORamos:2025uqs,Laine:2025rll}, and the predictions presented in the literature exhibit discrepancies.
Throughout this work we use three methods WI1, WI2 and WI3. Method WI1 is essentially a slight refinement of the approach taken in \cite{Mirbabayi:2022cbt}. }
	\label{WIcomputations}
\end{table}

\begin{table}
\centering
\begin{tabular}{c  c | c c c c c}
benchmark & method & $\x$ & FWHM$_\x$ & masss &  $\log_{10}\frac{\Treh}{\rm GeV}$ \\
\hline
\ref{it:A} & WI1 & $12.1\pm 0.6$  & 0.33 & 31\% & $13.9\pm0.6$\\
\ref{it:A} & WI2 & $12.5\pm0.6$ & 0.09 & 37\% & $14.3\pm0.6$\\
\ref{it:C} & WI1 & $12.5\pm0.6$ & 0.16 & 50\% & $14.3\pm0.6$\\
\ref{it:C} & WI2 & $12.7\pm0.5$ & 0.08 & 56\% & $14.5\pm0.5$\\
\end{tabular}
\caption{Parameters characterizing the posteriors displayed in Fig.~\ref{fig:WIpost}
for the warm inflation model \eqref{eq:lag}. Here \emph{method} refers to the different computational approaches in the references given in Table \ref{WIcomputations}.
\label{WIparameters}}
\end{table}

In the model  of QCD-driven warm inflation with a $\lambda\varphi^4$-potential \eqref{eq:lag} there is only one unconstrained combination of the parameters $\lambda$ and $f$ 
after imposing the current measurement of $A_s$, defining a line in the $n_s$-$r$ plane. 
Fig.~\ref{fig:warmellipses} shows that
a measurement of $r$ by AliCPT-1  could rule out a significant fraction of this line, 
imposing both an upper and lower bound on its extension in the $n_s$-$r$ plane.
Assuming a rapid transition from inflation to radiation domination, a measurement of $\Treh$ can then be extracted from \eqref{Twarm}.

\begin{figure}
	\centering
	\includegraphics[width=0.9\linewidth]{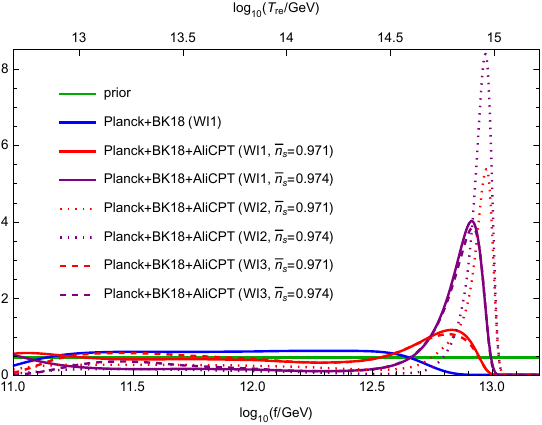}
	\caption{
    Posterior distributions for $\Treh$ and $f$ for Planck+BK and the observational benchmarks \ref{it:A} and \ref{it:C}, computed with the three methods introduced in the references given in Table \ref{WIcomputations}. 
    } 
	\label{fig:WIpost}
\end{figure}

However, a precise quantification of the knowledge gain 
on $f$, $\lambda$ and $\Treh$ requires reliable theoretical computations of the CMB anisotropies for given model parameters. Currently the theoretical uncertainties are comparable to the observational ones, cf.~Fig.~\ref{fig:warmellipses}. In  Fig.~\ref{fig:WIpost} we compare posterior distributions 
based on our principal  observational benchmark \ref{it:A} 
for the three different computational methods 
listed in Table \ref{WIcomputations}.
Additionally we consider an alternative observational benchmark \ref{it:C}, cf.~Table \ref{PrincipalBenchmark}. 
While the conclusion that AliCPT-1  can considerably improve the constraints on $\Treh$ as well as the microphysical parameters $\lambda$ and $f$ is true for both observational benchmarks and all three computational methods, quantifying this knowledge gain in case of an observation will require improved theoretical computations of the CMB spectra for given model parameters. 

Independently of the computational method the posteriors exhibit a pronounced peak near the observational upper limit on $\Treh$ and a tail towards smaller values. 
The reason for the appearance of this tail is that the generation of thermal fluctuations is very efficient in the strong regime of warm inflation, leading to a steep suppression of $r$ when increasing $f$. This in turn leads to a very mild dependence of $f$ on $r$, implying that the exponential suppression of the likelihood \eqref{Eq:Likelihood} is not efficient enough to cut the tail. 
As a result, the posterior variance $\sigma_{\log_{10}\lambda}$ after adding AliCPT-1  data does not improve much compared to Planck+BK, 
in spite of the fact that the posteriors in Fig.~\ref{fig:WIpost} visually clearly indicate a knowledge gain.
In the best case (benchmark \ref{it:C} with method WI2) it improves from $\sigma_\x\simeq0.6$ to  $\sigma_\x\simeq0.5$, in the worst case (benchmark \ref{it:A} with method WI1) it actually grows from $\sigma_\x \simeq 0.5$ to  $\sigma_\x\simeq 0.6$),
with $\x=\log_{10}(f/{\rm GeV})$.

Given their highly non-Gaussian shape, the variance is not the best way to quantify this knowledge gain. A possibly more suitable indicator is the full width at half maximum (FWHM) of the posterior peak, which in term of $\log_{10}\lambda$ improves from $5.5$ to 0.4, with a peak weight of 59\%  for benchmark \ref{it:C} with method WI2. Even for 
benchmark \ref{it:A} with method WI1 there is still an improvement from $5.1$ to $1.1$ with a peak weight of 37\%. With regard to the constraints on $\log_{10}(f/\text{GeV})$, for benchmark \ref{it:C} with method WI2 
the FWHM improves from $1.8$ to $0.08$ with a peak weight of 56\%, while for benchmark \ref{it:A} with method WI1 
the FWHM improves from $1.7$ to $0.33$ with a peak weight of 31\%.  
 For benchmark \ref{it:C} with method WI2, the improved variance and FWHM for $\log_{10}(\Treh/\text{GeV})$ are $0.5$ and $0.03$, respectively, compared to $0.6$ and $0.10$ for benchmark \ref{it:A} with method WI1.

\section{Discussion and Conclusion}

AliCPT is the only major CMB observatory planned in the Northern Hemisphere in foreseeable time. One of its main science goals is the detection of primordial gravitational waves through a measurement of the scalar-to-tensor ratio $r$. This makes AliCPT sensitive to the reheating phase after inflation, the impact of which on the CMB can in first approximation be parameterised in terms of $r$ and the spectral index $n_s$. 
Assuming a fiducial value of $r=0.01$, we showed that 5 years of AliCPT-1  
observations with 19 modules can considerably improve our knowledge of cosmic reheating. 

In plateau $\alpha$-attractor models with fixed $\alpha$ this could, under favorable conditions, amount to a simultaneous measurement of the scale of inflation $\M$
with an accuracy around $1\%$
and the order of magnitude of the
reheating temperature 
with an accuracy slightly worse than $10\%$,
cf.~Table \ref{ErrorBars}. 
If reheating is driven by a Yukawa coupling, this can be translated into a measurement of the coupling constant by means of \eqref{YukawaConstraint}. 
If $\alpha$ is simultaneously fitted, neither $\alpha$ nor $\Treh$ can be measured individually.
This is a consequence of the challenge \ref{challenge1} that most models of inflation  that are consistent with data face: the potential contains too many free parameters to measure all of them.
However, the constraint on the combination of the two parameters $\M$ and $\alpha$ can improve considerably compared to current observations.  This could rule out about half of the currently allowed interval for $\alpha$ and increase the lower bound on $\Treh$ by up to two orders of magnitude, see Fig.~\ref{fig:2DpostalphaT}.
The situation is qualitatively similar in RGI models, but the constraints would be slightly weaker for the fiducial values considered, 
and no lower bound on $\Treh$ can be derived when simultaneously fitting $\alpha$, see Fig.~\ref{fig:2DpostRGI}.  

In QCD-driven warm inflation with a quartic potential there are only two parameters, both of which can be constrained by adding AliCPT data to current observations.
In this case two of the main challenges in testing theories of inflation can be addressed with AliCPT:
\ref{challenge1} the only parameter $\lambda$ in the inflationary potential can be constrained and \ref{challenge2} the inflaton coupling to SM particles $f$ can be measured, with no need for any hypothetical new particles as mediators during reheating.
Hence, the cosmic history could be fully reconstructed. Moreover, since the inflaton can in principle be found in axion searches, a measurement of $f$ by AliCPT would lead to a prediction that could be tested in the laboratory in the future. However, currently the accuracy at which $f$ can be determined is limited by theoretical uncertainties, as indicated by the comparison of computational methods in Fig.~\ref{fig:WIpost}.

Our results demonstrate AliCPT's  potential to gain information on $\Treh$ and the inflaton coupling. 
The former is of great interest for both particle physics and cosmology because it corresponds to the initial temperature of the hot big bang, the latter is the microphysical parameter connecting cosmic inflation and particle physics. 
Hence, AliCPT can make an important contribution to addressing several important questions in cosmology and fundamental physics: 
\ref{it:expansion}  the expansion history before the epoch of radiation domination,
\ref{it:thermal}  the thermal history during reheating and
\ref{it:micro}  the connection between models of cosmic inflation and particle physics theories within which these are embedded.
The AliCPT receiver was successfully deployed at the observatory and achieved first light in early 2025. This crucial milestone marks its transition from construction to scientific operation. The next phase aims to deploy the full focal plane detector array within five years, and integrating over 32,000 detectors will enable AliCPT to enter its ultimate observation stage and fully achieve its designed detection capability.
Our findings add weight to the science case for AliCPT, as it can help to reveal the first moments of cosmic history and provide clues to understand the connection between cosmic inflation and fundamental physics.

\vspace{0.5cm}
\textbf{Acknowledgments.}
We are grateful to Prof.~Xinmin Zhang for his support and for helping to make this collaboration possible.
This work has been partially funded by the National Natural Science Foundation of China No. 12403005, the National Key R\&D Program of China No. 2020YFC2201601 and by the Deutsche Forschungsgemeinschaft (DFG, German Research Foundation) - SFB 1258 - 283604770.

\begin{appendix}

\section{Relation to CMB observables 
}\label{sec:appendixA}
In standard single-field cold inflation scenarios, 
$\Nreh$ can be related to 
the number of $e$-folds $\Nk$ between the horizon-crossing of a cosmological perturbation with wave number $k$ and the end of inflation via
\begin{equation}
	\label{Nre}
	\begin{split}
		N_{\rm re} &= \frac{4}{3\Bar{w}_{\rm re}-1}\bigg[\Nk+\ln\left(\frac{k}{a_0 T_0}\right)+\frac{1}{4}\ln\left(\frac{40}{\pi^2g_*}\right)\\
		&\quad +\frac{1}{3}\ln\left(\frac{11g_{s*}}{43}\right)-\frac{1}{2}\ln\left(\frac{\pi^2M^2_{ pl}r A_s}{2\sqrt{\Vend}}\right)\bigg],
	\end{split}
\end{equation}
where $a_0$ and $T_0=2.725~{\rm K}$ are the scale factor and the temperature of the CMB at the present time, respectively, 
and we approximate $g_{s*} = g_*$.
$\Nk$ is found from
\begin{equation}\label{Nk}
	\Nk=\ln\left(\frac{a_{\rm end}}{a_k}\right)=\int_{\varphi_k}^{\varphi_{\rm end}}\frac{H d\varphi}{\dot{\varphi}}
	\approx\frac{1}{M^2_{pl}}\int_{\varphi_{\rm end}}^{\varphi_k}d\varphi\frac{\V }{\partial_\varphi \V }~.
\end{equation}
Here $H_k, \varphi_k$ etc.~denote the value of  
$H, \varphi$ etc.~at the time when the pivot-scale $k$ crosses the horizon. The value of $\varphi_k$ in terms of 
$n_s$ and $r$ is found 
by solving
\begin{equation}
	n_s=1-6\epsilon_k+2\eta_k~, \quad r=16\epsilon_k.
	\label{nANDr}    
\end{equation}
In the slow roll regime $\epsilon,\eta \ll 1$, we find 
\begin{equation}
	\label{H_k}
	H^2_k=\frac{\V(\varphi_k)}{3M_{pl}^2}~
	=\pi^2 M_{pl}^2\frac{r A_s}{2}.
\end{equation}
This permits to express $\Treh$ in terms of the observables $\{A_s, n_s, r\}$
by inserting \eqref{Nre} with \eqref{Nk} into \eqref{Tre}, 
where $\varphi_k$ is determined by solving \eqref{nANDr} for $\varphi_k$, and $\Vend$ as well as $\varphi_{\rm end}$ can be determined by solving the condition $\epsilon=1$ for $\varphi$.
From \eqref{nANDr} one obtains
\begin{equation}
	\epsilon_k=\frac{r}{16}~,\quad \eta_k=\frac{n_s-1+3r/8}{2},
\end{equation}
which, together with the definitions of $\epsilon$ and $\eta$, yields
\begin{equation}\label{TakaTukaUltras}
\frac{\partial_\varphi \V}{\V}\Bigl|_{\varphi_k}=\sqrt{\frac{r}{8M_{pl}^2}} \ , \quad
\frac{\partial^2_\varphi \V}{\V}\Bigl|_{\varphi_k}=\frac{n_s-1+3r/8}{2M_{pl}^2}.
\end{equation}
Together \eqref{TakaTukaUltras} and \eqref{H_k} can be used to relate the effective potential and its derivatives to $\{A_s, n_s, r\}$, and to express $\wrehbar$ and $\Nreh$ in \eqref{Nre} in terms of these observables, allowing to express the RHS of \eqref{GammaConstraint} in terms of $(n_s,A_s,r)$ \footnote{We note in passing that the procedure presented here assumes that quantum corrections to the inflaton trajectory are negligible, which has recently been put into question for $\alpha$-attractor models \cite{Alexandre:2025ixz}.}.

\section{Measuring the inflaton coupling}\label{sec:appendixB}

In standard cold inflation, the expression \eqref{CouplinfToTreh} can be used to extract the inflaton coupling from \eqref{GammaConstraint} if \eqref{GammaPerturbative} holds when $\GG=H$.
However, in general there is backreaction in the form of feedback effects of the produced particles on $\GG$  (including non-perturbative effects and resonant particle production \cite{Amin:2014eta}) that invalidate the use of the simple expression \eqref{GammaPerturbative}. 
This does not only pose a practical computational challenge, but also represents a conceptual limitation because the feedback effects can be very sensitive to the interactions of the produced particles with each other (scatterings, further decay chains), which depend on their properties (masses, couplings, etc.). 
Hence, $\GG$ in general depends on a large number of parameters of the underlying particle physics theory.
This makes it impossible to impose a meaningful constraint on any individual parameter from the limited set of observables $\{A_s, n_s, r\}$. 

Suppose that reheating is driven by an interaction between $\varphi$ and a set of other fields $\{\mathcal{X}_i\}$ of the form $\g \varphi^\n \EFTscale^{4-\DD}\mathcal{O}[\{\mathcal{X}_i\}]$. 
Here $\mathcal{O}[\{\mathcal{X}_i\}]$ is an operator of mass dimension $\DD-\n$
and $\g$ a dimensionless coupling constant. 
Scalar interactions, Yukawa couplings and axion-like couplings all fall into this category.
A conservative criterion for neglecting the impact of feedback on the expansion history is given by \cite{Drewes:2019rxn}
\begin{equation}
	|\g| \ll\left(\frac{m_\varphi}{\varphi_{\rm end}}\right)^{\n-\frac{1}{2}}
	{\rm min}\left(
	\sqrt{\epsilonvarM}
	,
	\sqrt{\epsilonvar}
	\right)
	\left(\frac{m_\varphi}{\EFTscale}\right)^{4-\DD}. \label{GeneralScaling}
\end{equation}
Avoiding feedback effects from the production of inflaton particles out of the oscillating condensate (sometimes called fragmentation) imposes a similar bound on the non-linear coefficients in the potential \footnote{The condition \eqref{GeneralScalingSelf} is derived under the assumption that the effective  single field description holds during reheating \cite{Drewes:2019rxn}. It has been pointed out that multi-field effects can play an important role in $\alpha$-T models \cite{Krajewski:2018moi}, but our values for $\alpha$ are considerably larger than those considered there and in the follow-up works \cite{Iarygina:2018kee,Iarygina:2020dwe}.},
\begin{equation}
	|\vv_\n| \ll \left(\frac{m_\varphi}{\varphi_{\rm end}}\right)^{\n-\frac{5
		}{2}}
	{\rm min}\left(
	\sqrt{\epsilonvarM}
	,
	\sqrt{\epsilonvar}
	\right)
	\left(\frac{m_\varphi}{\EFTscale}\right)^{4-\n}.\label{GeneralScalingSelf}
\end{equation}
In plateau-potentials this approximately translates into 
\begin{eqnarray}\label{YippieYaYaySchweinebacke}
	|\vv_\n| \ll \left(3\pi^2  r A_s\right)^{(\n-2)/2} , \ 
	|\g| \ll \left(3\pi^2  r A_s\right)^{\n/2}.
\end{eqnarray}
The conditions \eqref{YippieYaYaySchweinebacke} have been derived based on simple analytic considerations in \cite{Drewes:2019rxn} and were found to be consistent with lattice simulations in \cite{Lozanov:2017hjm,Garcia:2021iag}.
If taken at face value, they would restrict the size of the inflaton coupling to be comparable to the electron Yukawa coupling in the SM or smaller.

However, regarding the \emph{expansion history} (time evolution of the scale factor $a$) of the universe and the \emph{thermal history} (time evolution of the temperature $T$) one can distinguish three regimes.
\begin{enumerate}[(i)]
	\item \label{it:reg1}
    For very small inflaton coupling $\g$, Hubble expansion can assure that the number densities of the produced particles remain sufficiently low at all times that feedback effects can be neglected. Then \eqref{GammaPerturbative} holds in good approximation during the entire reheating epoch. The thermal history and expansion history can both be obtained by solving the equations of motion 
	for the averaged energy densities $\dot{\rho}_\varphi+3H\rho_\varphi+\GG \rho_\varphi =0 $ 
	and $\dot{\rho}_R+4H\rho_R-\GG \rho_\varphi=0$ along with the Friedmann equation \cite{Martin:2010kz}. This set of equations can be solved analytically to estimate both the maximal temperature during reheating $T_{\rm max}$ and the reheating temperature $\Treh$ \cite{Giudice:2000ex}, with the latter given by
	\begin{eqnarray}
		T_{\rm re}=\sqrt{\GG M_{pl} }\left(\frac{90}{\pi^2g_*}\right)^{1/4}\Big|_{\GG=H}.\label{TRsimple}
	\end{eqnarray}
	In this regime, observational constraints on $\Nreh$ can be translated into bounds on the inflaton coupling $\g$ by inserting \eqref{GammaPerturbative} into \eqref{GammaConstraint}, and with this knowledge the thermal history during reheating can be fully reconstructed. 
	\item \label{it:reg2} 
    For intermediate values of $\g$, feedback effects modify $\GG$ during the early stage of reheating, but do not alter the moment when $\GG=H$. The thermal history is changed 
    \footnote{If the feedback is purely thermal \cite{Drewes:2013iaa}, then the coupled equations for $\rho_\varphi$ and $\rho_R$ can still be solved analytically \cite{Drewes:2014pfa,Co:2020xaf,Ming:2021eut} and $T_{\rm max}$ be determined if one assumes instant thermalisation of the produced particles. While this assumption is reasonable during most of the reheating period \cite{Harigaya:2013vwa}, it is questionable during the quick initial rise of $\rho_R$ \cite{Mukaida:2015ria,Garcia:2021iag}, possibly leading to an overestimate of $\T_{\rm max}$.
    }, 
	which can have an important impact on the production of relics \cite{Garcia:2021iag,Ai:2023qnr}, but the expansion history remains unaffected, hence there is no impact on CMB observables. 
	$\Treh$ is still given by \eqref{TRsimple}, and observational constraints on $\g$ can still be obtained by inserting \eqref{GammaPerturbative} into \eqref{GammaConstraint}. However, this knowledge is insufficient to fully reconstruct the thermal history during reheating at early times, when it is affected by feedback.
	\item \label{it:reg3} 
    For large $\g$, feedback modifies both the thermal history and expansion history. In this case it is generally not possible to impose a constraint on $\g$ or any other individual microphysical parameter from the CMB, as the $\GG$ in general depends on a large number of parameter that affect the interactions of the produced particles with each other \footnote{If the inflaton directly couples to the SM and no other mediator particles with unknown properties play a significant role during reheating (such as in Higgs inflation \cite{Bezrukov:2007ep} or QCD-driven warm inflation \cite{Berghaus:2025dqi}), then the relation between $\g$ and $\Nreh$ is calculable unambiguously even in regime \ref{it:reg3}. In that case one can still
		extract information on the inflaton coupling from the CMB even if the conditions
		\eqref{GeneralScaling} and \eqref{GeneralScalingSelf} are violated, see Fig.~\ref{fig:WIpost}.}. 
\end{enumerate}
A determination of the reheating temperature $\Treh$ from \eqref{Tre} is possible in all three regimes \ref{it:reg1}-\ref{it:reg3}. 
A determination of the inflaton coupling from \eqref{CouplinfToTreh} is only possible in regimes \ref{it:reg1} and \ref{it:reg2}.
Using this result to reconstruct the full thermal history during reheating is only possible in regime \ref{it:reg1} and requires the additional assumption that the dissipation has been dominated by the same interaction throughout the whole reheating epoch.

The conditions \eqref{GeneralScaling} and \eqref{GeneralScalingSelf} are in general too conservative to be used as a criterion that marks the transition between regimes \ref{it:reg2} and \ref{it:reg3}.    
In particular, for a Yukawa coupling to fermions, Pauli blocking tends to suppress resonant particle production, so that $\Treh$ is determined by perturbative decays \cite{Giudice:1999fb,Berges:2010zv,Fan:2021otj,Garcia:2021iag,Xu:2025wjq}.
The Pauli blocking itself is also a form of feedback, and assuming that the occupation numbers are not significantly reduced  (e.g.~by decays and scatterings  \footnote{\label{foot:Pauli}
In particular, the effective mass $\sim y\varphi$ that the Yukawa interaction generates makes the fermions very heavy during maximal inflaton elongations $\varphi \sim \varphi_{\rm end}$, implying that they might decay between two zero crossings of $\varphi$, and that the Pauli blocking that renders the non-perturbative production of fermions inefficient is avoided.
Other effects that may justify the validity of \eqref{YukawaConstraint} beyond \eqref{GeneralScaling} are even more model dependent (for instance, thermal fermion masses could reduce the efficiency of non-perturbative particle production).  
To remain fully agnostic about the produced particles' unknown interactions,  the conditions \eqref{GeneralScaling} and \eqref{GeneralScalingSelf} should be applied.}) in principle brings in a model dependence.
However, on general grounds one can expect that a reduction of the occupation numbers will extend the validity of the perturbative result \eqref{GammaPerturbative}. 
Hence, applying \eqref{YukawaConstraint} throughout  our analysis only amounts to mild model assumptions

The results for the Yukawa coupling can in principle be translated in a straightforward manner to other types of interactions. 
For instance, a scalar coupling $g\varphi\chi^2$ 
or an axion-like coupling $\frac{\upalpha}{8\pi f}\varphi F_{\mu\nu}\tilde{F}^{\mu\nu}$
correspond to choosing $(\g,c)=(g/m_\varphi,8\pi)$ or $(\g,c)=(\upalpha m_\varphi/(8\pi f),4\pi)$, respectively. However, for interactions with bosons the transition from regime \ref{it:reg2} to \ref{it:reg3} generally occurs for smaller values of $\g$, hence feedback effects considerably limit the range of values for $\g$ in which the use of \eqref{CouplinfToTreh} can be justified.

\end{appendix}

\bibliography{bib.bib}

\end{document}